\documentclass[showpacs,amsmath,amssymb,floatfix,prl,twocolumn]{revtex4}
\usepackage[pdftex]{graphicx}
\usepackage{amssymb,amsfonts,amsmath}
\usepackage{graphicx,amsmath}

\newcommand{\beq}{\begin{eqnarray}}
\newcommand{\eeq}{\end{eqnarray}}

\newcommand{\dd}{\mathrm{d}}

\newcommand{\unt}[1]{\,\mathrm{ #1} }

\newcommand{\prt}[1]{\left(#1\right) }
\newcommand{\cro}[1]{\left[#1\right] }

\newcommand{\tah}[1]{\tanh\!\left[#1\right] }
\newcommand{\ch}[1]{\cosh\!\left[#1\right] }
\newcommand{\sh}[1]{\sinh\!\left[#1\right] }

\newcommand{\abs}[1]{\left\lvert#1\right\rvert}
\newcommand{\norme}[1]{\left\lVert#1\right\rVert}

\begin{document}

\title{Transport properties of overheated electrons trapped on a Helium surface}
\author{F. Closa}
\affiliation{Laboratoire de Physico-Chimie Th\'eorique, UMR CNRS Gulliver 7083, ESPCI, Paris France}
\affiliation{Laboratoire de Physique Théorique de la Matière Condensée, UMR 7600 LPTMC 4, place Jussieu, 75252 Paris Cedex 05, France}
\author{E. Rapha\"el}
\affiliation{Laboratoire de Physico-Chimie Th\'eorique, UMR CNRS Gulliver 7083, ESPCI, Paris France}
\author{A.D. Chepelianskii}
\affiliation{Cavendish Laboratory, University of Cambridge, J J Thomson Avenue, Cambridge CB3 OHE, UK}

\pacs{89.20.Hh, 89.75.Hc, 05.40.Fb} 

\begin{abstract}
An ultra-strong photovoltaic effect has recently been reported for electrons 
trapped on a liquid Helium surface under a microwave excitation tuned at intersubband resonance 
[D. Konstantinov et. al. : J. Phys. Soc. Jpn. {\bf 81}, 093601 (2012) ].
In this article, we analyze theoretically the redistribution of the electron density
induced by an overheating of the surface electrons under irradiation, and obtain quantitative 
predictions for the photocurrent dependence on the effective electron temperature 
and confinement voltages. We show that the photo-current can change sign
as a function of the parameters of the electrostatic confinement potential on the surface,
while the photocurrent measurements reported so far have been 
performed only at a fixed confinement potential. The experimental observation of this sign reversal
could provide a reliable estimation of the electron effective temperature
in this new out of equilibrium state. Finally, we have also considered the effect of the 
temperature on the outcome of capacitive transport measurement techniques. 
These investigations led us to develop, numerical and analytical methods for solving 
the Poisson-Boltzmann equation in the limit of very low temperatures which could 
be useful for other systems.
\end{abstract}

\maketitle

Electrons trapped on the liquid helium surface were the first experimental realization 
of a high mobility two dimensional conductor \cite{andrei,monarkha_book}. The extremely high 
mobility of the surface electrons allows to explore in an unique way many problems in fundamental physics, 
examples include Wigner crystallization \cite{WignerCystal}, propagation of magnetoplasmon waves \cite{glatli,volkov}, 
transport of correlated charges in confined geometries \cite{drees,harkov}, 
quantum melting \cite{melting} and sliding \cite{sliding}.
The interest in this system has been
renewed by its potential for quantum computation that comes from 
their very large spin coherence times \cite{Schuster} 
and from the spacial addressability of the electrons on the surface \cite{Mukharsky,Mika}. 
Recently, an unexpected transport regime was observed 
in novel experiments where surface electrons were excited by a Millimeter-wave irradiation 
\cite{Konstantinov2009,Konstantinov2010}.
This regime occurs when the perpendicular magnetic field lies within regularly spaced intervals 
for which the ratio between the microwave and cyclotron frequencies is close to an integer value.
It also appears only when the photon energy is equal to the energy spacing between the two lowest transport subbands.
This transition is usually called inter-subband resonance, 
it corresponds to transitions between the ground state and the first excited state of the electronic wavefunction
in the direction perpendicular to the Helium surface.
Once the above conditions on the perpendicular magnetic field and on the photon-energy are satisfied, 
the irradiation can lead to a complete suppression of the dissipative component 
of the electronic response in capacitive measurements. This effect is strikingly similar 
to zero-resistance states that were first reported in ultra high mobility GaAlAs hetero-structures \cite{Mani2002,Zudov2003}.

Further experiments on surface electrons showed a strong redistribution of the 
electronic density compared to its equilibrium shape under zero-resistance conditions \cite{DenisAlexei}. 
This redistribution was characterized by a depletion of the electronic
density at the center of the electron cloud and by an accumulation of electrons towards the edge of the cloud.
One of the motivations for these experiments was the possibility of electron 
trapping at the edges of the system 
under microwave irradiation that has been proposed theoretically \cite{Toulouse}
and confirmed recently in extensive numerical simulations \cite{Toulouse2}.
However other mechanisms could also be responsible for this redistribution. 
Indeed, a theoretical description of the magneto-resistance oscillations under irradiation 
has recently been proposed by Y.P. Monarkha \cite{Monarkha1,Monarkha2}.
Although this theory does not directly predict a redistribution of the electronic density similar 
to that observed in the experiments, it has been suggested that such a redistribution could occur,
due to the instability of the negative resistance state that appears in this theory \cite{IvanReview}. 
Finally a strong overheating of the surface electrons could also result in a significant 
redistribution of the electronic density. 
In order to distinguish between these different mechanisms that compete with each other
we analyze theoretically the overheating scenario in the present article.
We show that in this case the direction 
of the photo-current can be reversed by changing the confinement gate potentials,
a behavior which is not expected for other redistribution mechanisms
and that could thus be important to determine the out-of equilibrium electron temperature.

When surface electrons (SE) absorb microwave irradiation
their effective temperature can increase significantly above 
the temperature of the Helium bath. 
At zero magnetic field, the effective temperature of the 
electrons under microwave pumping 
of the intersubband resonance was studied both theoretically 
and experimentally in \cite{Denis2007}. It was shown 
that electrons could be overheated to temperatures in the 
range of $T_e \simeq 10\;{\rm K}$ even if the temperature of the 
Helium bath was much lower $T_0 \simeq 0.3\;{\rm K}$,
however no estimation of $T_e$ is available in the zero resistance state (ZRS) regime.
Experimentally it was shown that the formation of ZRS is accompanied by 
a strong redistribution of the electrons on the Helium surface, 
which can cost up to a $100\;{\rm meV}$ charging energy per electron \cite{DenisAlexei}.
Thus the electrons need to absorb many microwave photons (typical frequencies are in the $100\;{\rm GHz}$ range)
to store the needed energy, which suggests that they can occupy 
highly excited states. It is probable that their energy 
distribution cannot be described appropriately 
by an effective temperature in this regime. However, since this 
is the simplest starting model to understand the distribution of the 
electrons under irradiation, we have investigated the effect 
of a high electron temperature on the electron density profile 
$n_e(\mathbf{r})$ in the trapping potential. As the charging energy cost is surprisingly 
high we are led to consider temperatures much larger than the thermal energy of the bath,
we note however that these temperatures are still an order of magnitude lower than 
the penetration barrier into the liquid Helium or than the height of the confining potentials 
which are both in the $eV$ range. 

\section{I. Simulation of hot electron density distribution}

\begin{figure}
\includegraphics[width=8cm]{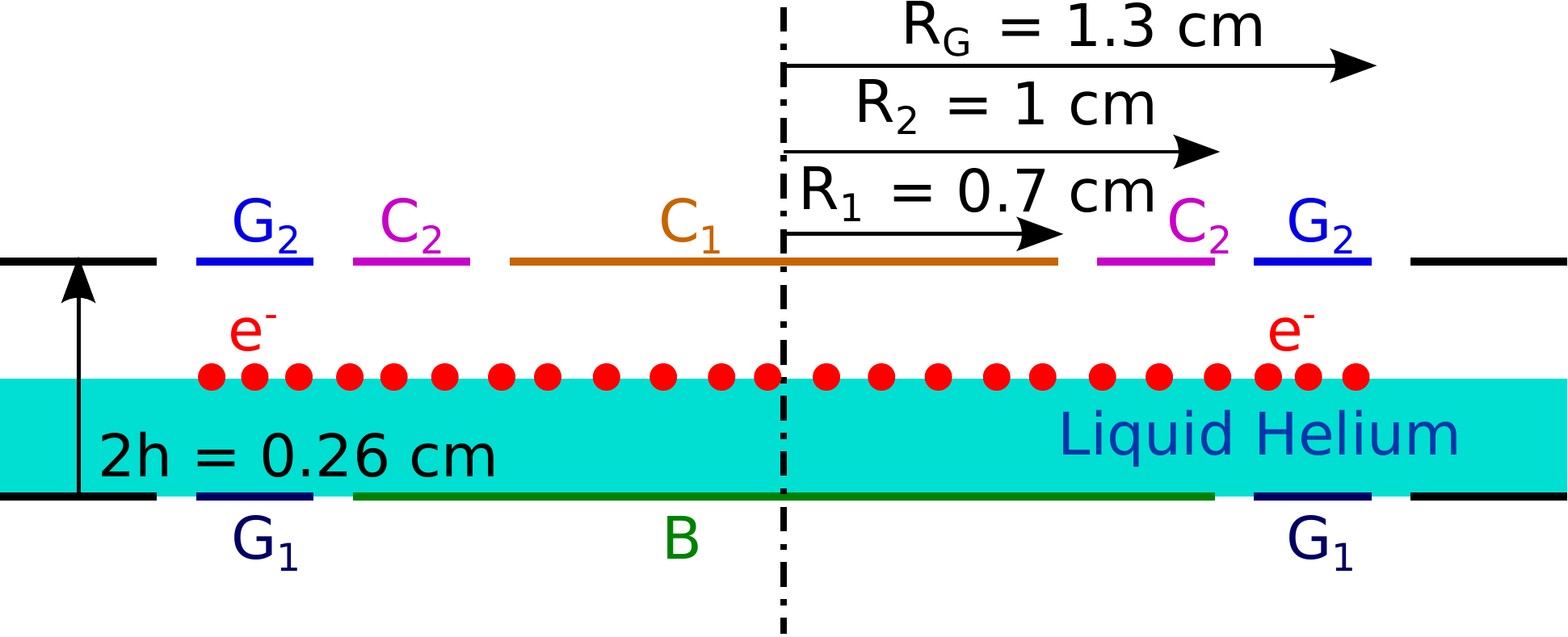}
 \caption{Schematic diagram of the experimental cell modeled in our simulations.
The cell is cylindrically symmetric, 
surface electrons are trapped above the central disc electrode $B$ and 
the bottom guard electrode $G_1$. The potential of $B$ was fixed to $V_B = 4.2\;{\rm V}$ 
in all the simulations (a typical experimental value). Electrodes $C_2$ and $C_1$ are grounded, 
except in section IV where a small AC potential is applied to $C_1$. 
In all simulations the potential difference between $G_1$ and $G_2$ is fixed to $V_B$ for reasons
that are explained in the text. The voltage $V_{G1}$ is thus the main control parameter 
for the shape of the electron cloud. 
} 
\label{FigCell}
\end{figure}

Since SE form a non degenerate system, their distribution on the
surface is governed by a Boltzmann statistics at 
the electron temperature $T_e$: 
\begin{align}
n_e(\mathbf{r}) = \frac{N_e}{
\int d^2 \mathbf{r'} \exp\left( \frac{e V(\mathbf{r'})}{k_B T_e} \right) } 
\exp\left( \frac{e V(\mathbf{r})}{k_B T_e} \right)
\label{eq:boltzmann}
\end{align}
Here $e$ is the absolute value of the electron charge, $V(\mathbf{r})$ the electrostatic potential created by the confining electrodes and the electron cloud. 
The quantity $N_e$ is the total number of electrons trapped in the cloud.
The electrostatic potential $V(\mathbf{r})$ is determined by solving 
the Poisson equation in a cylindrical cell that is sketched on Fig.~\ref{FigCell}. 
The potential and electron density depend only on the distance $r$ to the cell axis and the potential $V(\mathbf{r})$ can be expressed 
in the following form:
\begin{align}
V(r) = V_{ext}(r) + \int G(r, r') n_e(r') 2 \pi r' d r'
\label{eq:poisson}
\end{align}
where $V_{ext}(r)$ is the potential created by the confining electrodes alone,
in absence of the electron cloud, and $G(r, r')$ a Green-function 
giving the potential created by electrons located at radius $r'$. 
Analytic formulas for $V_{ext}(r)$ and the Green-function are derived in the appendix 
for the case where the effect of the finite radius of the experimental cell can be neglected
(note also that the electrostatics of surface electrons has been analyzed in \cite{Wilen,Kovdrya}). 
This approximation is accurate in the limit where 
the difference between the radius of the 
experimental cell and the radius of the electron cloud is larger than the 
cell height $2 h$ (see Fig.~\ref{FigCell}). Finite elements simulations in a more realistic
geometry confirmed that finite cell size effects are expected to be negligible 
in the experimental geometry used in \cite{Konstantinov2009,Konstantinov2010,DenisAlexei}.
We also assumed that the electron density distribution remains two dimensional 
even for hot electrons which is accurate as long as the electron temperature remains 
much smaller than the potential difference between top and bottom electrodes.
Indeed in this limit the thickness of the electron cloud is still negligible 
before the height of the cell $2 h$.
The appendix gives an explicit analytic formula for the potential 
 $V_{ext}(r)$ as a function of the potential of the bottom disc $B$ and guard 
electrodes $G_{1,2}$ under the assumptions that were indicated above. 
The combination of Eqs.~(\ref{eq:boltzmann},\ref{eq:poisson}), forms 
a Poisson-Boltzmann equation that needs to be solved in order to determine 
the electron density $n_e(r)$ and the electrostatic potential $V(r)$. 
Interestingly we have previously studied a similar problem in the 
context of counterion condensation around permeable hydrophobic globules \cite{Globule}.

The numerical solution of this equation in the limit of low 
(but finite) temperatures is challenging since small errors on the 
value of the potential $V(r)$ can lead to a large errors on the electron 
density $n_e(r)$ as a consequence of Eq.~(\ref{eq:boltzmann}).
This can introduce instabilities in many numerical procedures
(for example direct iteration of Eqs.~(\ref{eq:boltzmann},\ref{eq:poisson}) 
or relaxation methods based on drift-diffusion equations). 
From our numerical experiments, we found that the best stability
at low temperatures was achieved by a Monte-Carlo computational
method. 

In this method, an ensemble of $N_t \simeq 10^4$ trial particles
were displaced on the simulated region of the Helium surface, 
using a Metropolis algorithm in the potential landscape 
given by Eq.~(\ref{eq:poisson}). 
The unknown density $n_e(r)$ that appears 
in this equation was determined by 
averaging over time the 
probability of presence of the trial particles in 
thin circulars shells in which the surface 
of the electron cloud was separated 
(typically the available configuration space was divided into $100$ 
circular shells). The outlined numerical method was stable for temperatures 
as low as $T_e = 0.1\;{\rm K}$, convergence was checked by computing the 
relative error: 
\begin{align}
\epsilon_{err} = \frac{1}{N_e} \int \left| n_e(r) - \frac{N_e}{Z} 
\exp\left( \frac{e V(\mathbf{r})}{k_B T_e} \right) \right| d^2 \mathbf{r}
\end{align}
with $Z = \int d^2 \mathbf{r'} \exp\left( \frac{e V(\mathbf{r'})}{k_B T_e} \right)$ the normalization factor in the Boltzmann-distribution. In this expression the potential $V(\mathbf{r})$ was computed 
by numerical integration of Eq.~(\ref{eq:poisson}) for the electron density obtained from the Monte-Carlo algorithm.
Note that a more detailed description of the numerical method that we used here will be published elsewhere \cite{Fabien2013}. 
The value of this relative error parameter in our simulations was around $\epsilon_{err} \simeq 10^{-3}$ at high temperatures ($T_e \simeq 100\;{\rm K}$), $\epsilon_{err} \simeq 5\times 10^{-3}$ at $T_e \simeq 1\;{\rm K}$ and $\epsilon_{err} \simeq 4 \times 10^{-2}$ at the lowest temperatures $T_e \simeq 0.1\;{\rm K}$;
as discussed above $\epsilon_{err}$ is higher at low temperatures due to the longer convergence times. 

\begin{figure}
\includegraphics[width=8cm]{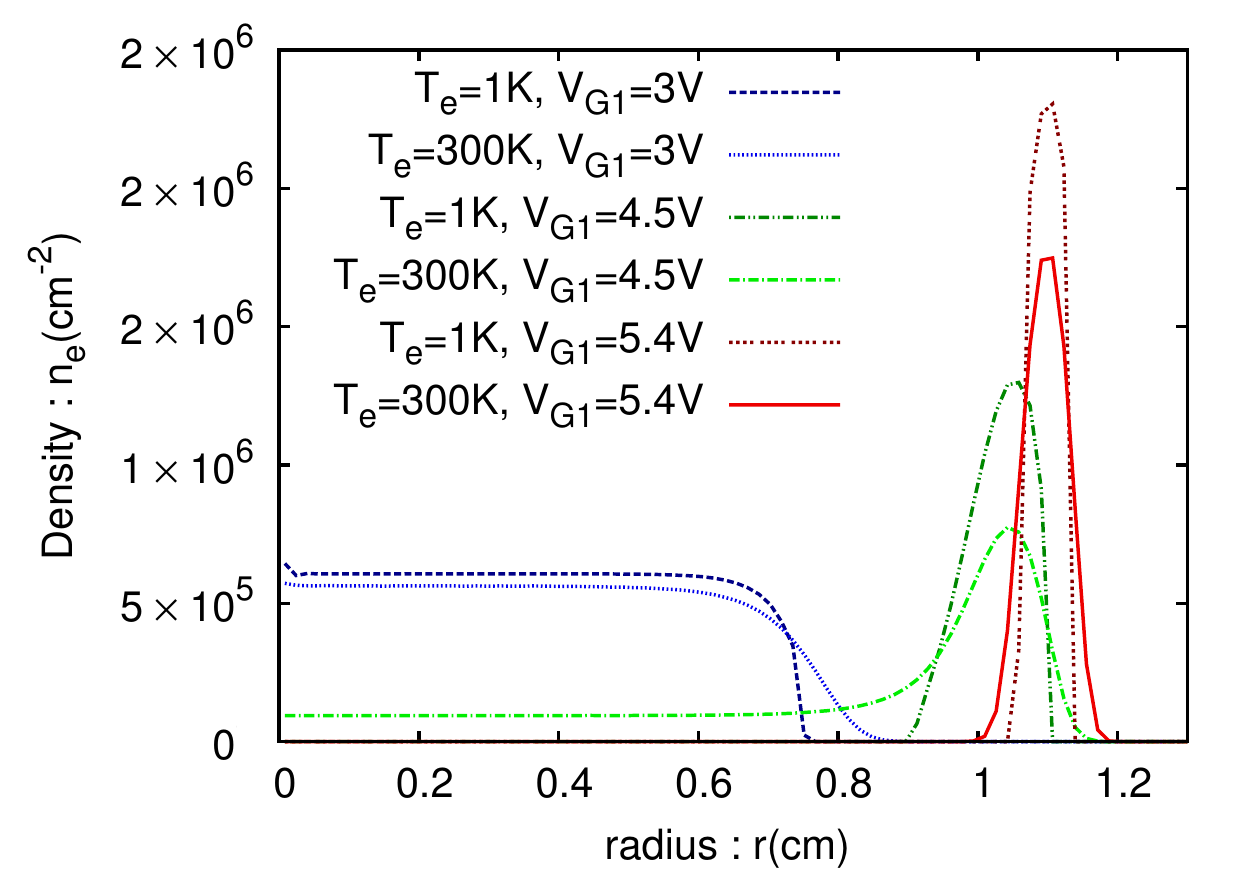}
 \caption{ 
Electronic Density profiles $n_e\prt{r}$ at the liquid helium surface as a function of the radius $r$ from the center of the cell, represented at two different electron temperatures ($T_e = 1\unt{K}$ and $T_e = 500 \unt{K}$) and for different potentials of the electrode $G_1$ ($V_{G1} = 3 \unt{V} $ : blue curves ; $5\unt{V}$ : green curves and $5.4\unt{V}$ : red curves) with a fixed total number of electrons $N_e = 10^6$. The potential of the lower central electrode $B$ is $V_B = 4.2 \unt{V}$. 
} 
\label{FigDensity}
\end{figure}

Typical electron density distributions are shown in Fig.~\ref{FigDensity} for several values of the gate 
parameters for $N_e = 10^6$ electrons in the cloud. As the potential in the guard electrodes is increased 
we observe a transition between different shapes of the electron cloud. 
In the limit of a low potential on the guard electrodes, the electrons are confined in the central region of the cell
below the electrode $C_1$ and the shape of the electron cloud is described by a monotonic distribution where the density vanishes at large radius $r$. 
This case corresponds to the curves at $V_{G1} = 3\;{\rm V}$ which are represented on Fig.~\ref{FigDensity}.
The value of the bias voltage between the guard $G_1$ and the disc $B$ is then $V_{G1} - V_{B} = -1.2\;{\rm V}$.
As in the experiments, a potential is also applied to the top guard electrode $G_2$ in order to enforce the relation  
$V_{G1} - V_{G2} = V_B = 4.2\;{\rm V}$. This constraint
allows to keep the perpendicular component of the electric field constant 
across the cell which is important to keep the same intersubband transition energy for all
the electrons in the system. 
The second type of electron configuration occurs when the potential of the peripheral electrode $V_{G1}$ is much higher than $V_{B}$,
the electrons are then confined only above the guard electrode $G_1$, forming a ring (see the curves at $V_{G1} = 5.4 \unt{V}$ on Fig.~\ref{FigDensity}). 
Finally an intermediate case appears when the Coulomb repulsion between electrons or their thermal energy is 
sufficiently strong to overcome the energy barrier created by the positive bias potential $V_{G1} - V_{B}$.
In this case the electrons are mainly confined under the electrode $G_2$ but a part of the electrons  
density is also localized above the central electrode $B$ forming a non-monotonic disk shaped distribution $n_e(r)$.  
This case is represented on Fig.~\ref{FigDensity} for $V_{G1} = 4.5 \unt{V}$ and $T_e = 1\unt{K}$ or $300\unt{K}$. Note the transition between a ring shaped density to a non-monotonic disk which is observed when temperature 
increases at $V_{G1} = 4.5 \unt{V}$.

The comparison between low and high temperature curves for these three cases reveals the following trend.
In the case of a monotonic ring density the electron cloud tends to expand outwards as the electrons are heated,
however when electrons form a non-monotonic disc density the trend is reversed and the electrons tend to 
come back towards the center as $T_e$ increases. This effect can be qualitatively understood as the tendency
of the electron temperature to smooth the electron density profile in the cloud.

\begin{figure}
\includegraphics[width=9cm]{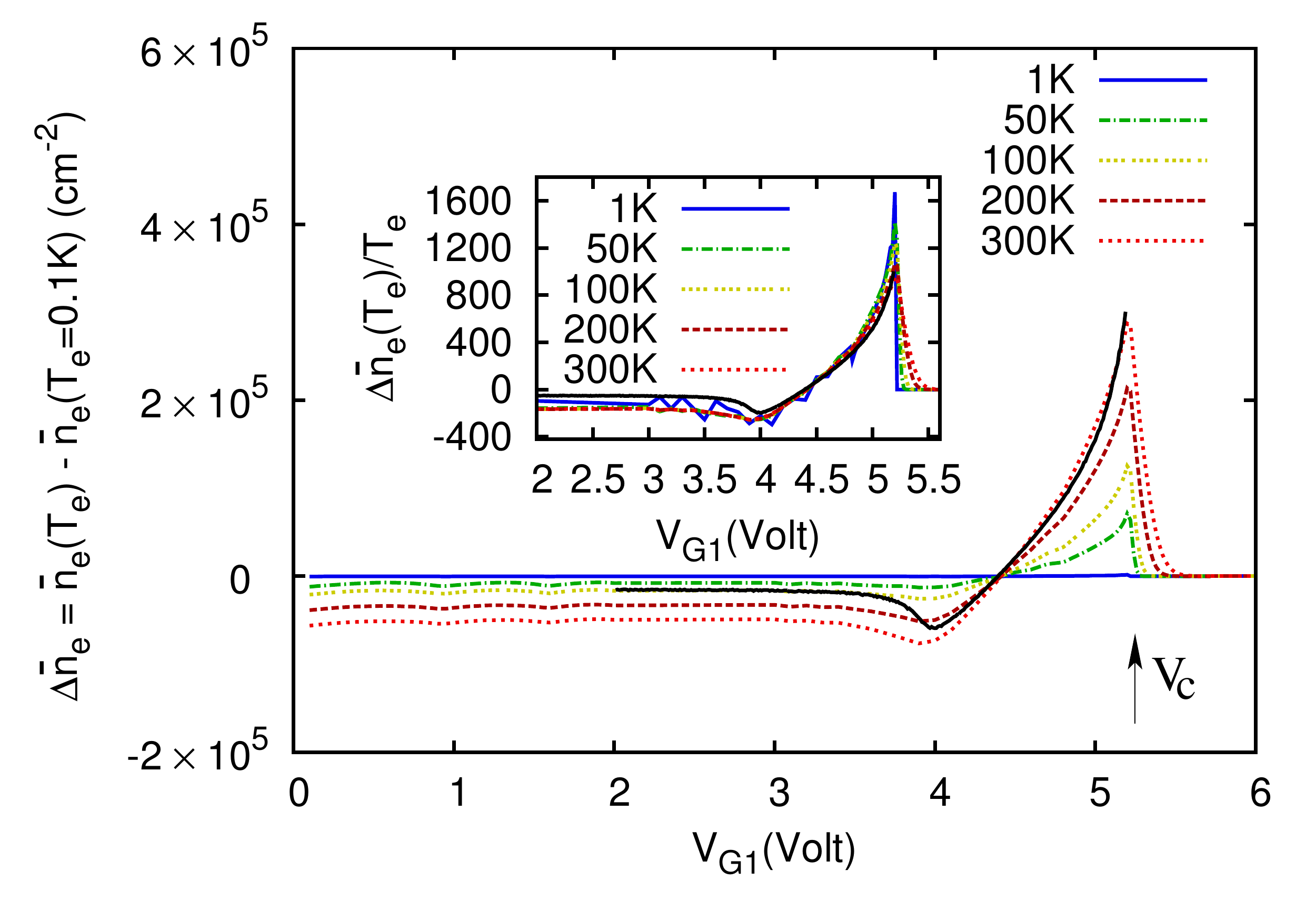} 
 \caption{ Electronic density variation $\Delta {\bar n}_e = \cro{{\bar n}\prt{T_e} - {\bar n}\prt{T_e=0.1 \unt{K}}}$ as a function of the guard potential $V_{G1}$ for different effective temperatures (from top to bottom at $V_{G1} = 1\;{\rm Volt}$: $0.1\unt{K}$, 
$50\unt{K}$, $100\unt{K}$, $200\unt{K}$ and $300\unt{K}$). 
The plain black curve corresponds to the predictions of the analytic theory
Eq.~(\ref{eq:deltane},\ref{eq:nav}) at effective temperature $T_e = 300 \unt{K}$.
In the inset, the density variation is rescaled by the temperature. For $V_{G1} < V_{C}$ 
all curves collapse on each other with a very good agreement with the analytical predictions.  
For $V_{G1} > V_{C}$, the scaling breaks down since the photocurrent becomes thermally activated (see discussion in the text). The number of electrons in the cloud was fixed to: $N_e = 10^7$.
} 
\label{Fig:Diff}
\end{figure}

In order to study quantitatively the effects of electron heating on the electron density 
we have focused on a quantity that can be accessed directly in experiments. 
In \cite{DenisAlexei} the change of the electron density was probed by measuring the 
transient currents created on the top electrodes $C_1, C_2, G_1$ by a periodic 
change of the density profile induced by an On/Off modulation of the microwave power 
under ZRS conditions. Since the electrodes $C_1$ and $C_2$ are held at the same potential
$n_e(r)$ is almost constant below these two electrodes, thus a simple 
plane capacitor model can be used to relate the transient current $i_1(t)$ measured by a current amplifier 
connected to $C_1$ and the change in the SE density:
\begin{align}
i_1(t) = -\frac{e}{2} \frac{d}{d t} \int_{S(C_1)} n_e(\mathbf{r}) d^2 \mathbf{r}
\end{align}
where $S(C_1)$ denotes the Helium surface below the electrode $C_1$. 
Within the frame of our simplified model, the On/Off modulation of the microwave power 
changes the electron temperature between its equilibrium value $T_0 \simeq 100\;{\rm mK}$ 
and its (unknown but possibly several order of magnitude larger) steady state value $T_e$ under illumination.
The time-integral of the photo-current during a microwave ON half-cycle will thus be given by:
\begin{align}
\int i_1(t) dt = -\frac{e}{2} S(C_1) ( {\bar n}_e(T_e) - {\bar n}_e(T_0) )
\label{eq:photocurrent}
\end{align}
where ${\bar n}_e$ denotes the average electron density below the central electrode $C_1$.
Thus knowing $\Delta {\bar n}_e = {\bar n}_e(T_e) - {\bar n}_e(T_0)$, 
it is possible to estimate the amount of charge displaced in the experiment.
The dependence of $\Delta {\bar n}_e$ on the guard voltage 
is displayed on Fig.~\ref{Fig:Diff} for several temperatures $T_e$.

This figure shows that $\Delta {\bar n}_e$ changes sign as a function of the guard voltage $V_{G1}$, 
and exhibits a sharp maximum for $V_{G1}$ close to the transition from a ring to a non-homogeneous disc which 
occurs at $V_{G1} = V_C$ with $V_C \simeq 5.21\;{\rm Volt}$, this value is obtained from numerical simulations at low 
temperatures. 
For potential $V_{G1} < V_B$, a temperature increase causes an 
expansion of the cloud below the electrode $V_{G1}$ leading to $\Delta {\bar n}_e < 0$. 
The quantity $\Delta {\bar n}_e$ changes sign at $V_{G1} \simeq V_B = 4.2\;{\rm Volt}$
(more precisely at a slightly higher value $V_{G1} \simeq 4.45\;{\rm Volt}$) and 
when the bias voltage $V_{G1} - V_B$ becomes positive we observe $\Delta {\bar n}_e > 0$.
We note also that the density change $\Delta {\bar n}_e$ almost vanishes when
SE form a ring with a vanishing density in the center at $V_{G1} > V_C$.
The latter behavior can be explained with the following argument: 
at zero temperature when $V_{G1} > V_{C}$ the potential at the center of the cell 
is lower than the potential inside the electron ring. This potential 
difference creates an energy barrier that precludes the electron from exploring the center of the cell 
when the thermal energy is small compared to the barrier height.
The maximal values of $\Delta {\bar n}_e$ are obtained when the electron cloud is in a non-homogeneous
disc configuration $V_B < V_{G1} < V_{C}$. This is intuitively plausible since in this case the center of the cell 
and the ring are at the same potential at $T_e = 0$ and there is no energy barrier to prevent hot electrons 
from exploring the center of the cloud. As expected $\Delta {\bar n}_e$ increases with temperature.

The above heuristic arguments can explain the sign of $\Delta {\bar n}_e$ as a function of $V_{G1}$,
but fail to account for some surprising aspects of the functional dependence such as the seemingly non-analytic 
behavior at $V_{G1} \simeq V_C$. Indeed $\Delta {\bar n}_e(V_{G1})$ features a sharp maximum for $V_{G1} < V_{C}$,
but quickly vanishes at $V_{G1} > V_{C}$, the asymmetry of this peak becomes even more pronounced at lower $T_e$. 
In the next section, we develop a perturbation theory that allows to explain this singularity 
and its temperature dependence.

\section{II. Perturbative calculation of the effect of temperature}

In this section, we derive a simplified analytic theory to explain the behavior of the SE when the electronic 
temperature is raised from $0$ to $T_e$. 
At zero temperature, the potential of the electron cloud is fixed and equal to a constant value $V_0$.
This allows to effectively linearize the equations since the electron cloud can be replaced by an equipotential surface. 
The boundaries of the cloud are then fixed by the constrain that the electric field vanishes on the boundary 
(stability condition) and the potential of the cloud $V_0$ is fixed by the number of electrons in the cloud $N_e$. 
Thus the problem of finding the zero temperature density distribution $n_e(r, T=0)$ is computationally 
much simpler, and we assume that $n_e(r, T=0)$ is known in the following derivation. 

If the temperature changes, the number of electron inside the cloud changes by an amount of $\delta n_e$, modifying the potential inside the cloud by an amount $\delta V$. We assume that the density inside the cloud is just a little perturbed $\delta n_e \ll n_e\!\prt{r, T_e=0}$:
\beq
n_e\prt{r,T_e} &=& n_e\!\prt{r, T_e=0} + \delta n_e(r).
\eeq
Using the Boltzmann-equation Eq.~(\ref{eq:boltzmann}) we also have:
\beq
n_e\prt{r,T_e}&=& n'_{av} \exp\prt{\dfrac{e V_0}{k_B T_e}} \exp\prt{\dfrac{e \delta V}{k_B T_e}}\nonumber\\
&=& n_{av} \exp\prt{\dfrac{e \delta V}{k_B T_e}},
\eeq
where $n_{av}$ and $n_{av}'$ are normalization constants. Using our assumption $\delta n_e \ll n_e\!\prt{r, T_e=0}$ we are led to
\beq
n_e\!\prt{r, T_e=0} &\simeq& n_{av} \exp\prt{\dfrac{e \delta V}{k_B T_e}}
\eeq
or equivalently: 
\beq\label{eq:deltaV}
e \delta V = k_B T_e \cro{\log n_e\prt{r, T = 0}-\log n_{av}}.
\eeq 
The quantity $\delta V$ can be estimated in an other way; in the center of the electron cloud where the density is uniform, the variation of density $\delta n_e$ leads to a change in the potential $\delta V$:
\beq\label{eq:capa}
\delta n_e = -\dfrac{2 \epsilon_0 \delta V}{e  h},
\eeq
this expression just comes from the electrostatics of a planar capacitor. 
The value of $\delta n_e$ taken at the center of the cell is equal to $\Delta {\bar n}_e$,
eliminating $\delta V$ between the equations (\ref{eq:deltaV}) and (\ref{eq:capa}), 
we obtain the following relation 
\beq\label{eq:deltane}
\Delta {\bar n}_e = \dfrac{2 \epsilon_0 k_B T_e \cro{\log n_e\prt{r = 0, T = 0} - \log n_{av}}}{e^2 h}.
\eeq
This equation is accurate only in the center of the electron cloud where density gradients can be neglected. 
Outside the electron cloud, the density decays exponentially with the temperature and can be neglected. 
If we tentatively assume that Eq.~(\ref{eq:deltane}) is valid in the entire cloud, 
we can find $n_{av}$ from the requirement of the conservation of electron number:
$\int \delta n_e r \dd r =0$, leading to  
\begin{align}
% &\int \cro{\log n_e\prt{r, T_e = 0}- \log n_{av}} = 0, \mbox{  or  } \nonumber 
n_{av} = \dfrac{\int_{cloud} \log [ n_e\prt{r, T = 0} ] r \dd r }{\int_{cloud} r \dd r}.
\label{eq:nav}
\end{align}
The integral in this expression is performed for the electron density at $T = 0$ for which 
the limits of the cloud are well defined.

The combination of Eqs.~(\ref{eq:deltane},\ref{eq:nav}) allows to find the variation of the 
electron density at the center of the cell based on the knowledge of the electron density at $T = 0$,
they also imply that $\Delta {\bar n}_e$ scales linearly with the temperature $T_e$ (we assume $T_e$
is much larger than the Helium bath temperature $T_0$).
The above analysis is justified only for $V_G < V_C$ where the density $n_e\prt{r = 0, T = 0}$
is non vanishing, we have thus compared our semi-analytical theory 
with the simulations only in this domain. The results are displayed on Fig.~\ref{Fig:Diff}, 
and show that our semianalytical approach is very successful for $V_G \simeq V_C$ where 
it reproduces the numerical results with very good accuracy without any adjustable parameters.
The agreement is worse in the limit $V_G \ll V_B$ probably because our assumption that 
Eq.~(\ref{eq:deltane}) is valid in the entire cloud does not hold anymore when the electrons are strongly 
confined by the guard electrodes. 
The scaling $\Delta {\bar n}_e \propto T_e$ is also confirmed by the simulations as it is illustrated 
in the inset of Fig.~\ref{Fig:Diff}. 
To summarize the proposed theory seems to explains quantitatively the change of the electron density under heating,
and allows detailed comparison with numerical simulations and hopefully future experiments. 

\section{III. Effect of heating on transport measurements}

In previous sections we have investigated the effect of an elevated out of equilibrium temperature 
on the distribution of the electrons on the Helium surface. The information on electron density 
can be accessed through photo-current measurements similar to those performed in \cite{DenisAlexei}.
However the most wide-spread type of experiments on SE is a Sommer-Tanner technique which
allows to determine the electron longitudinal mobility $\mu_{xx}$ \cite{Sommer}. 
We have thus analyzed numerically the effect of $T_e$ on the outcome of such experiments.

In a Sommer-Tanner measurement an A.C. potential $V_{ac} \cos \omega t$ is applied to one of the top electrodes 
(we have taken this electrode to be $C_1$ in our simulations) an the induced AC 
displacement of image-charges is detected from the remaining ones using a lock-in technique.
Our numerical approach to simulate this experiment was a direct numerical integration of the time dependent 
drift-diffusion equations:
\begin{align}
\frac{\partial n_e}{\partial t} + \frac{\mu_{xx}}{r} \frac{\partial}{\partial r} \left( r n_e \frac{\partial V}{\partial r}   - \frac{k_B T}{e} r \frac{\partial n_e}{\partial r} \right) = 0
\label{eq:driftdif}
\end{align}
This equation is coupled to a Laplace equation on the electrostatic potential,
which includes the electron density and the static and time dependent potentials on the electrodes as
a boundary conditions. The drift-diffusion equations were integrated using an implicit first order 
integration scheme solved with a finite element method (FEM). For reference we give here the weak formulation of the 
differential equation connecting $n_e(r,t+\Delta)$ and $n_e(r,t)$ in our numerical calculations: 
\begin{align}
&\frac{1}{\mu_{xx}} \int \left[ n_e(r, t+\Delta t) - n_e(r, t) \right] u(r) r dr = \nonumber  \\  
&\int \left(n_e(r, t+\Delta t) \frac{\partial V}{\partial r} \frac{\partial u}{\partial r} - \frac{k_B T}{e} \frac{\partial n_e(r, t+\Delta t)}{\partial r} \frac{\partial u}{\partial r} \right) r dr 
\end{align}
the potentials are evaluated at time $t$ and recomputed at time $t + \Delta t$ ($\Delta t$ is our integration time step) 
using the updated value of the density $n_e$, $u(r)$ denotes a finite element trial function. 
Finite element simulations were performed using the FreeFem programming language \cite{FreeFem},
for typical parameters $N_e = 3 \times 10^6$, $V_{B} = 4.2\;{\rm V}$, $V_{G1} = V_{G2} = 0\;{\rm V}$. 
The AC potential with amplitude $V_{ac} = 3\;{\rm mV}$ was applied to the electrode $C_1$, 
and the modulation of the image-charge number 
was determined on the neighboring electrode $C_2$ after subtraction of the direct capacitive coupling. 
This modulation $Q(t)$ was separated into in-phase and out of phase components: $Q_x$ and $Q_y$ which 
respectively denote the amplitude of the $\cos \omega t$ and $\sin \omega t$ components in the total signal $Q(t)$.
An immediate consequence of Eq.~(\ref{eq:driftdif}) is that the quantities $Q_x$ and $Q_y$ depend only on 
the ratio $\frac{\omega}{\mu_{xx}}$, this is a consequence of the fact that the potentials react 
instantaneously to the displacement of charges at the typical experimental frequencies in the ${\rm kHz}$ range. 
This quantity should then be compared with the parameters of the electron cloud which, for 
a monotonic disc density distribution are the density at the cloud center $n_e$ and the cloud radius $R_c$. 
These considerations lead us to define the dimensionless parameter $W = \frac{\epsilon_0 \omega R_c}{\mu_{xx} e n_e}$ 
which characterizes the ratio between the frequency and the response time of SE.

\begin{figure}
\includegraphics[width=8cm]{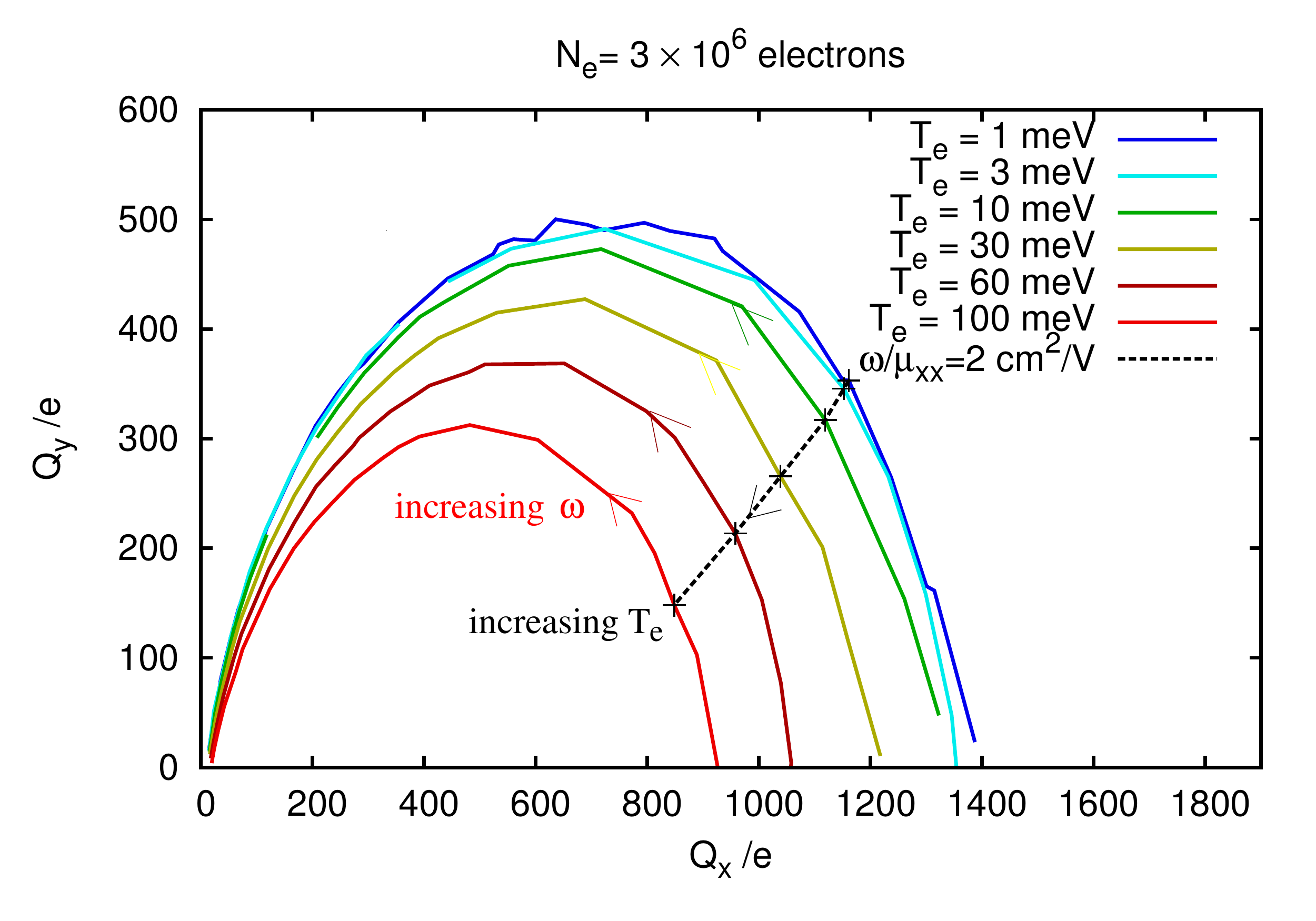}
 \caption{Cole-cole diagram of the in-phase ($Q_x$) and out of phase ($Q_y$) charge modulation on electrode $C_2$
as the modulation frequency $\omega$ is changed. The AC potential is applied to $C_1$ and 
the charge modulation due to the direct capacitive coupling between both electrodes was subtracted. 
The continuous curves were obtained at different temperatures that are indicated in the legend 
(as temperature increases the semicircle radius decreases), the direction of increasing $\omega$ (decreasing $\mu_{xx}$) is indicated by the arrows on the curves. The dashed curve with symbols joins
points at different temperatures for a fixed value of $\omega/\mu_{xx} = 2\;{\rm cm^2/V}$. 
} 
\label{Fig:AC}
\end{figure}

We summarized our numerical results in a Cole-Cole diagram that displays trajectories followed by $Q_x$ and $Q_y$ on $(Q_x,Q_y)$ plane 
as the measurement frequency $\omega$ is varied at fixed $T_e$. 
At low frequencies when $W \ll 1$ 
the electrons have time to follow the time varying potential adiabatically 
and all the displaced image-charge is in phase with the excitation $Q_x \gg Q_y$. As frequency increases, an out of phase 
component appears since the electrons are unable to respond fast enough to follow the phase of the external potential. 
Finally at frequencies much higher that the electron characteristic response time $W \gg 1$ 
their response vanishes and the trajectory converges to the origin $(Q_x = 0, Q_y = 0)$. 

The simulation results at different temperatures are shown on Fig.~\ref{Fig:AC}, 
the obtained trajectories in the $(Q_x, Q_y)$ plane are very similar to semicircles.
As the temperature is increased the radius of the semicircle tends to shrink, qualitatively this occurs because 
some of the terminally exited electrons are not sensitive to the applied bias potential. 
As a consequence the shape of the trajectories in the $(Q_x, Q_y)$ plane, are very different if the 
mobility of the electrons changes under microwave irradiation or their temperature. 
Indeed when $\mu_{xx}$ is changed the points $(Q_x, Q_y)$ still collapse on their 
semi-circular trajectory because it is parametrized only by the ratio $\frac{\omega}{\mu_{xx}}$. 
On the contrary, when the temperature changes the points $(Q_x, Q_y)$  lie on different semicircles 
which leads to very different trajectories. An example of a trajectory versus temperature 
at fixed $\omega/\mu_{xx} = 2 \;{\rm cm^2/V}$ is shown Fig.~\ref{Fig:AC}. Thus heating can
create deviations of the $(Q_x, Q_y)$ trajectories from the behavior that would be expected if only 
the electron mobility changed under microwave irradiation.

\section{IV. Conclusions}

We investigated the effect of an elevated out of equilibrium electron temperature on the distribution of surface electrons 
on liquid Helium in a typical experimental geometry. Our main result is probably the determination of the 
photocurrent induced by an increase of the 
electron temperature under irradiation. 
This was achieved numerically by developing a Monte-Carlo method 
capable of computing the electron density even in the limit of very low temperatures,
and analytically through a perturbation theory analysis of the effects of temperature.
The dependence of the photocurrent on the control gate voltage that is predicted within  
our model is quite specific and should provide a reliable signature of electron heating,
against other mechanisms possibly at play.
Namely we predict that the photocurrent generated by density variation at the center of the 
cloud changes sign as a function of the guard voltage 
and is stronger in the reverse bias configuration.
Unexpectedly, we also find that the photo-current is maximal when 
center of the electron cloud is almost depleted.
Finally we have also demonstrated numerically that in Sommer-Tanner experiments, 
heating should induce trajectories on the Cole-Cole diagram that differ strongly from those expected from a change 
in electron mobility. This could be checked in transient conductance measurements,
where the trajectory on the Cole-Cole plane can be measured as the system is driven in and out 
of a zero resistance state by an on/off modulation of the microwaves power.
We hope that our results will allow a quantitative analysis of the electron energy 
distribution under zero-resistance conditions, and that the analytical and numerical methods 
developed here will be useful in other systems that are described by a Poisson-Boltzmann equation.

We are thankful to E. Trizac, D. Konstantinov and K. Kono for fruitful discussions.
One of us, A.C. acknowledges the support from RIKEN and St Catharine college in Cambridge. 

\section{Appendix}

In this appendix, the electrode potential $V_{ext}\prt{\mathbf{r}}$ and the Green function $G\prt{\mathbf{r},\mathbf{r'}}$ are calculated. The total potential $V\prt{\mathbf{r}}$ is the sum of the electrode potential $V_{ext}\prt{\mathbf{r}}$ and the contribution coming from the electrons $V_e\prt{r} = \int G\prt{\mathbf{r},\mathbf{r'}} n_e\prt{\mathbf{r}} \dd^3 \mathbf{r}$. 
The cylindrical symmetry of the system allows us to write the different potential as functions of the cylindrical coordinate $\mathbf{r}=\prt{r, \theta , z}$ but we also make use the Eulerian coordinate $\mathbf{r}=\prt{x,y,z}$. Note that in our notations 
$\norme{\mathbf{r}} \ne r$.

\subsection{Electrode potential $V_{ext}\prt{\mathbf{r}}$}

The potential Electrode $V_{ext}\prt{\mathbf{r}}$ is solution of the Poisson Equation
\beq\label{poissonVext}
\Delta V_{ext} = 0. 
\eeq
Taking the $xy$-Fourier transform of this equation (\ref{poissonVext}), we get :
\beq\label{partialVext}
\partial_{zz} \tilde{V}_{ext} - k^2 \tilde{V}_{ext} = 0
\eeq
where $k^2 = k_x^2 + k_y^2$, $\mathbf{k} = \prt{k_x, k_y}$ and 
\beq
V_{ext}\prt{\mathbf{r}} = \iint \dfrac{\mathrm{d}^2\mathbf{k}}{2 \pi} e^{i \mathbf{k} \cdot \mathbf{r}} \tilde{V}_{ext}\prt{\mathbf{k},z}.
\eeq
The solution of Eq.(\ref{partialVext}) are 
\beq
\tilde{V}_{ext}\prt{k_x, k_y, z} = A(k_x, k_y)\cosh(k z) \nonumber \\ + B(k_x, k_y) \sinh(kz)
\eeq
where $A\prt{k_x, k_y}$ and $B\prt{k_x, k_y}$ are functions which satisfy the bordering conditions.
The cylindrical symmetry imposes $A\prt{k_x, k_y} = A\prt{k}$ and $B\prt{k_x, k_y} = B\prt{k}$, leading to 
\beq\label{PotVextTF}
V_{ext}\prt{r, \theta, z} = \int_0^{\infty} k \dd k  J_0\prt{k r} \left[A(k)\cosh(k z) \right. \nonumber \\ + \left. B(k) \sinh(kz) \right],
\eeq
where $J_i\prt{x}$ is the $i-th$ Bessel function of the first kind. Quoting $V_i$ the potential of the electrode $i$, $R_i$ the maximal radius of the electrode $i$ and $H\prt{r<R_i}$ the Heaviside function (equal $1$ if the argument is true, else zero), the electrode potential has to respect the following boundary conditions :
\begin{align}
&V_{ext}\!\prt{r, z= h}=V_{G_2} \cro{H\!\prt{r<R_{G_2}} - H\!\prt{r<R_{C_2}} } \label{TFCB2} \\  
&V_{ext}\!\prt{r, z= - h}=V_{B} H\!\prt{r<R_{B}} +\nonumber \\
& \;\;\;\;\;\;\;\;\;\;\;\;\;\;\;\;\; V_{G_1} \cro{H\!\prt{r<R_{G_1}} - H\!\prt{r<R_{B}}   }.
\label{TFCB}
\end{align}
It is straightforward to calculate 
\beq\label{TFH}
H\!\prt{r<R} = \int_0^{\infty} R J_1\!\prt{k R} J_0\prt{k r} \dd k.
\eeq
From the equations (\ref{TFCB2}), (\ref{TFCB}) and (\ref{TFH}), we obtain the following boundary conditions :
\begin{equation}\label{BC}
\begin{array}{l}
V_{ext}\!\prt{r, h}  \nonumber \\ =  \int_0^{\infty} J_0\!\prt{k r} V_{G_2} \cro{R_{G_2} J_1\!\prt{k R_{G_2}} - R_{C_2} J_1\!\prt{k R_{C_2}}} \dd k \vspace{0.1cm}\nonumber \\
=\int_0^{\infty} k J_0\!\prt{k r} \cro{ A\!\prt{k} \ch{k h} + B\!\prt{k} \sh{kh}} \dd k \\
%\end{array}
%\end{equation}
%and
%\begin{equation}\label{BC}
%\begin{array}{l}
V_{ext}\!\prt{r, -h}  \nonumber \\ 
=  \int_0^{\infty} J_0\!\prt{k r} \left[\prt{V_{B} - V_{G_1}} R_{B} J_1\!\prt{k R_{B}} \right.\nonumber \\
\hspace{3cm}\left.+ V_{G_1} R_{G_1} J_1\!\prt{k R_{G_1}} \right] \dd k \nonumber \\
= \int_0^{\infty} k J_0\!\prt{k r} \cro{ A\!\prt{k} \ch{k h} - B\!\prt{k} \sh{kh}} \dd k \vspace{0.2cm}\,.
\end{array}
\end{equation}
Which leads to:
\begin{widetext} 
\beq
A\!\prt{k}&=&\dfrac{1}{2 k \ch{kh}} \left[V_{G_2} \cro{R_{G_2} J_1\!\prt{k R_{G_2}} - R_{C_2} J_1\!\prt{k R_{C_2}}} + \prt{V_{B} - V_{G_1}} R_{B} J_1\!\prt{k R_{B}} + V_{G_1} R_{G_1} J_1\!\prt{k R_{G_1}} \right],\\
B\!\prt{k}&=&\dfrac{1}{2 k \sh{kh}} \left[V_{G_2} \cro{R_{G_2} J_1\!\prt{k R_{G_2}} - R_{C_2} J_1\!\prt{k R_{C_2}}} - \prt{V_{B} - V_{G_1}} R_{B} J_1\!\prt{k R_{B}} - V_{G_1} R_{G_1} J_1\!\prt{k R_{G_1}} \right].
\eeq
\end{widetext}

With these expressions of $A\prt{k}$ and $B\prt{k}$, the electrode potential $V_{ext}\prt{r}$ is completely determined. 

\subsection{Electron potential $V_{e}\prt{\mathbf{r}}$}

Lets us consider an electron in the center of the system, that means localized at the coordinate $(0,0,0)$. 
This electron creates a potential $V_e\prt{\mathbf{r}}$. This potential can be written 
\beq
V_e\prt{\mathbf{r}} = -\dfrac{e}{4 \pi \epsilon_0} \prt{\dfrac{1}{\norme{\mathbf{r}}} + \bar{V}}\,.
\eeq 
The term $1/\norme{\mathbf{r}}$ is coming from the solution of the Laplace equation for a punctual charge and the term $\bar{V}$, solution of the Laplace equation $\Delta \bar{V} = 0$, permits at the potential to satisfy the boundary conditions $V_e\prt{r,\theta,z = \pm h} = 0$. As previously, the potential $\bar{V}$ can be obtained as, see Eq.(\ref{PotVextTF}) :
\beq\label{PotVeTF}
\bar{V}\prt{r, z} = \int_0^{\infty} k \dd k  J_0\prt{k r} \left[\bar{A}(k)\cosh(k z) \right. \nonumber \\ + \left. \bar{B}(k) \sinh(kz) \right].
\eeq
The function $\bar{B}\prt{k}$ must be equal to $0$ due to the symmetry of the problem. 
Furthermore, we can calculate 
\beq\label{TF1surr}
\dfrac{1}{\norme{\mathbf{r}}} &=& \int k J_0\!\prt{k r} \dfrac{e^{-k \abs{z}}}{k} \dd k \,,
\eeq 
leading to the full electronic potential $V_e\prt{\mathbf{r}}$ :
\beq
V_e\!\prt{r, z} \!=\! \dfrac{- e}{4 \pi \epsilon_0}  \! \int_{0}^{\infty}   \!\!\!\!\!\!\!k J_0\!\prt{k r} \!\!\cro{\bar{A}\!\prt{k} \ch{k z} \!+\! \dfrac{e^{-k\abs{z}}}{k}}\!\! \dd k. \nonumber \\
\eeq
The boundary conditions $V_e\prt{r, z = \pm h} = 0$ lead to the following expression for $\bar{A}\prt{k}$ :
\beq
\bar{A}\!\prt{k} = -\dfrac{e^{-kh}}{k \ch{k h}}\,, 
\eeq
and then 
\beq
V_e\!\prt{\rho, z} = \! \dfrac{- e}{4 \pi \epsilon_0}\! \int_0^{\infty}\!\!\!\!\!\!\! J_0\!\prt{k \rho}\!\! \cro{e^{- k \abs{z}} \!- \!e^{-k h} \dfrac{\ch{k z}}{\ch{kh}}} \!\!\dd k.
\eeq

\subsection{The complete potential $V\prt{\mathbf{r}}$}

To obtain the complete potential due to all electrons and electrodes, we have to sum the electrode potential and the potential created by the surface electrons distributed with a density $n_e\prt{r}$ in the plan $z=0$:
\beq
V\!\prt{r, z} &=& V_{ext}\!\prt{r, z} \nonumber \\&+ &\!\! \int_0^{R_c} \!\!\!\!\int_0^{2 \pi}  \!\!\!\!\!V_e\!\prt{\norme{\mathbf{r} - \mathbf{r'}},z} n_e\!\prt{r'} r' \dd r' \dd \theta .
\eeq
Now, we have to note that the electrons can move only in the plan $z=0$ and, by this fact, the electrons feel a potential $V\!\prt{r, z=0} \equiv V\!\prt{r}$ defined by :
\beq\label{Vtotz0}
V\!\prt{r} \!\!&=& \!\! V_{ext}\!\prt{r, z=0} \nonumber \\
\!&-&\!\! \dfrac{e}{2 \epsilon_0}\int_{0}^{\infty}\!\!\!\!\!\! \dd k\tah{kh}\!\! \int_{0}^{R_c}\!\!\!\!\!\!\dd r' r' J_0\!\prt{k r} J_0\!\prt{k r'} n\!\prt{r'} .\nonumber \\
\eeq
The comparison between Eq.(\ref{Vtotz0}) and Eq.(\ref{eq:poisson}) give the following expression for the Green function  
\beq
G\prt{r,r'} = \dfrac{- e}{4 \pi \epsilon_0} \int_0^{\infty} \dd k \tanh\cro{k h} J_0\!\prt{kr} J_0\!\prt{k r'} .
\eeq
The full potential $V\!\prt{r}$, see eq. (\ref{Vtotz0}), is numerically obtained. In order to increase the numerical precision, the integrations are done successively between two consecutive zeros of the Bessel functions $J_0\!\prt{kr}$ or $J_0\!\prt{k r'}$.

\clearpage 

\end{document}